\documentclass[aps,prl,10pt,twocolumn,psfig,superscriptaddress]{revtex4-2}
\usepackage{graphicx}
\usepackage{dcolumn}
\usepackage{bm}
\usepackage{amsmath}
\usepackage{lmodern}
\usepackage{amsmath}
\usepackage{color}
\usepackage{amssymb}
\usepackage{braket}
\usepackage{mathtools}
\usepackage{comment}
\usepackage[colorlinks=true,linkcolor=blue,urlcolor=blue,citecolor=blue]{hyperref}

\usepackage{ulem}

\newcommand{\dwave}[0]{{\it d}-wave }

\def\Fig#1{Fig.~\ref{#1}}

\begin{document}
\title{Robust d-wave superconductivity from the Su-Schrieffer-Heeger-Hubbard model:\\
possible route to high-temperature superconductivity}

\author{Hao-Xin Wang}
\affiliation{Institute for Advanced Study, Tsinghua University, Beijing 100084, China}
\author{Yi-Fan Jiang}
\thanks{jiangyf2@shanghaitech.edu.cn}
\affiliation{School of Physical Science and Technology, ShanghaiTech University, Shanghai 201210, China}
\author{Hong Yao}
\thanks{yaohong@tsinghua.edu.cn}
\affiliation{Institute for Advanced Study, Tsinghua University, Beijing 100084, China}
\date{\today}

\begin{abstract}
Increasing numerical studies showed that the simplest Hubbard model on the square lattice with strong repulsion may not exhibit high-temperature superconductivity (SC).
It is desired to look for other possible microscopic mechanism of realizing high-temperature SC.
Here, we explore the interplay between the Su-Schrieffer-Heeger (SSH) electron-phonon coupling (EPC) and the Hubbard repulsion by density-matrix-renormalization-group (DMRG) simulations. Our state-of-the-art DMRG study showed convincingly that the interplay between strong Hubbard $U$ and moderate Su-Schrieffer-Heeger EPC $\lambda$ can induce robust $d$-wave SC.
The SSH-type EPC can generate effective antiferromagnetic spin-exchange interactions between neighboring sites, which plays a crucial role in the interplay of inducing robust $d$-wave SC.
Specifically, for $U=8t$, we find that $d$-wave SC emerges when $\lambda>\lambda_c$ with a moderate critical value $\lambda_c=0.1\sim 0.2$.
Our results might shed new light to understanding high-temperature SC in cuprates as well as pave a possible new route in looking for high-temperature SC in other quantum materials with both strong $U$ and moderate $\lambda$.
\end{abstract}
\maketitle

{\bf Introduction:}
Since the experimental discovery of high-temperature superconductivity (SC) in cuprates nearly four decades ago, consensus about the microscopic mechanism for $d$-wave high-temperature SC has yet to be reached, which remains a key challenge in studies of strongly-correlated electronic systems \cite{Kivelson-Review2015, Davis2013, Dagotto1994, PALee-RMP2006, Scalapino-RMP2012, Fradkin2015}.
So far the probably most studied model for understanding SC in cuprates is the Hubbard model \cite{Anderson1987, Zhang1988}, which was believed by many to be the simplest possible model for producing both antiferromagnetic order and $d$-wave SC, depending on doping \cite{Kivelson-Review2015, Davis2013, Dagotto1994, PALee-RMP2006, Scalapino-RMP2012, Fradkin2015, Anderson1987, Zhang1988,Anderson-2004} (for a recent review, see Refs. \cite{Arovas-ARCMP2022,MPQin2022ARCMP}).
However, in the past several years, growing numerical studies \cite{SimonsCollaboration2015, SimonsCollaboration2017, Huang2017, Huang2018, SimonsCollaboration2020, Sorella-arXiv2021} suggest that the simplest Hubbard model with strong Hubbard interactions in hole doping range relevant for hole-doped cuprates exhibits stripe order \cite{Zaanen-PRB1989, Schultz-PRB1989, Kivelson-RMP2003} or  phase separation \cite{Kivelson-HQLin-PRL1990} but no $d$-wave SC in the ground state, opposite to conclusions of high-temperature SC ground state obtained from many earlier studies. 

It is natural to ask what more is needed beyond the simplest Hubbard model with strong $U$ in order to realize $d$-wave SC. Various works have attempted to address this issue.
For instance, by including the next-nearest-neighbor hopping $t'$ ($t'<0$ for cuprates), $d$-wave SC was shown to emerge for electron doping but whether $d$-wave SC emerges or not for hole doping remains elusive \cite{HongChen2019Science, Jiang2020, Shoushu2021, White2021PNAS, Peng2022, DNSheng2023, SWZhang2024Science};
other related studies include doping the $J_1$-$J_2$ spin liquid \cite{HongChen2021DopedJ1J2} or explicitly imposing stripe order in a deformed Hubbard model \cite{HongChen2022PNAS}.
Here, we propose to explore the interplay between the Su-Schrieffer-Heeger (SSH) electron-phonon coupling (EPC) and the Hubbard repulsion for realizing $d$-wave SC mainly for two reasons.
Firstly, increasing experimental and theoretical studies showed that in addition to electron-electron interactions EPC is crucial in understanding essential properties of various correlated quantum materials, including high-temperature SC in cuprates \cite{  Hirsch1987, Assaad1996,Assaad1997, Clay2020PRR, ZXShen2003RMP, Lanzara2001,ZXShen2002, Nagaosa2004B1gPhonon, EPC_APRES_Review2005, tallon2005, ZXShen2005, rosch2005, Lee2006, Reznik2006, ZXShen2021_1Dcuprate, ZXShen2018, he2018, nagaosa2004, gadermaier2010, devereaux2010, Liu2016, Nowadnick2012, Johnston2013, Sankar2014, Nowadnick2015, Greitemann2015, Nath2015, Nath2016, Costa2018, Costa2020, Wang2020,QKXue-NSR2021} and in Fe-based superconductors \cite{Wang_2012, Lee2014, DHLee-2015, ZXShen2017, Song2019, zhangshuyuan2019, weiwei2018, ruipeng2020, dennis2017, dunghai2018_Review, wang2016FeSe, zhouyuanjun2017, zixiang2019, zixiang2016, XHChen-NSR2014}.
It is the cooperative effect of EPC and electronic correlation that are deemed to be essential.
For instance, it was shown recently that the interplay between Holstein-type EPC and strong Hubbard $U$ may drive exotic quantum states such as pair-density-wave SC \cite{hongyao2020Holstein, Huang2022}.
Secondly, recently Ref. \cite{XunCai2021} (also see Refs. \cite{Assaad2022PRB,Scalettar2022PRB,XCai2022PRB}) made a numerical discovery that the SSH-type EPC on the square lattice at half-filling can induce AFM ordering, which usually helps realize unconventional SC upon doping \cite{Scalapino-RMP2012}.

\begin{figure}[t]
    \includegraphics[width=0.44\textwidth]{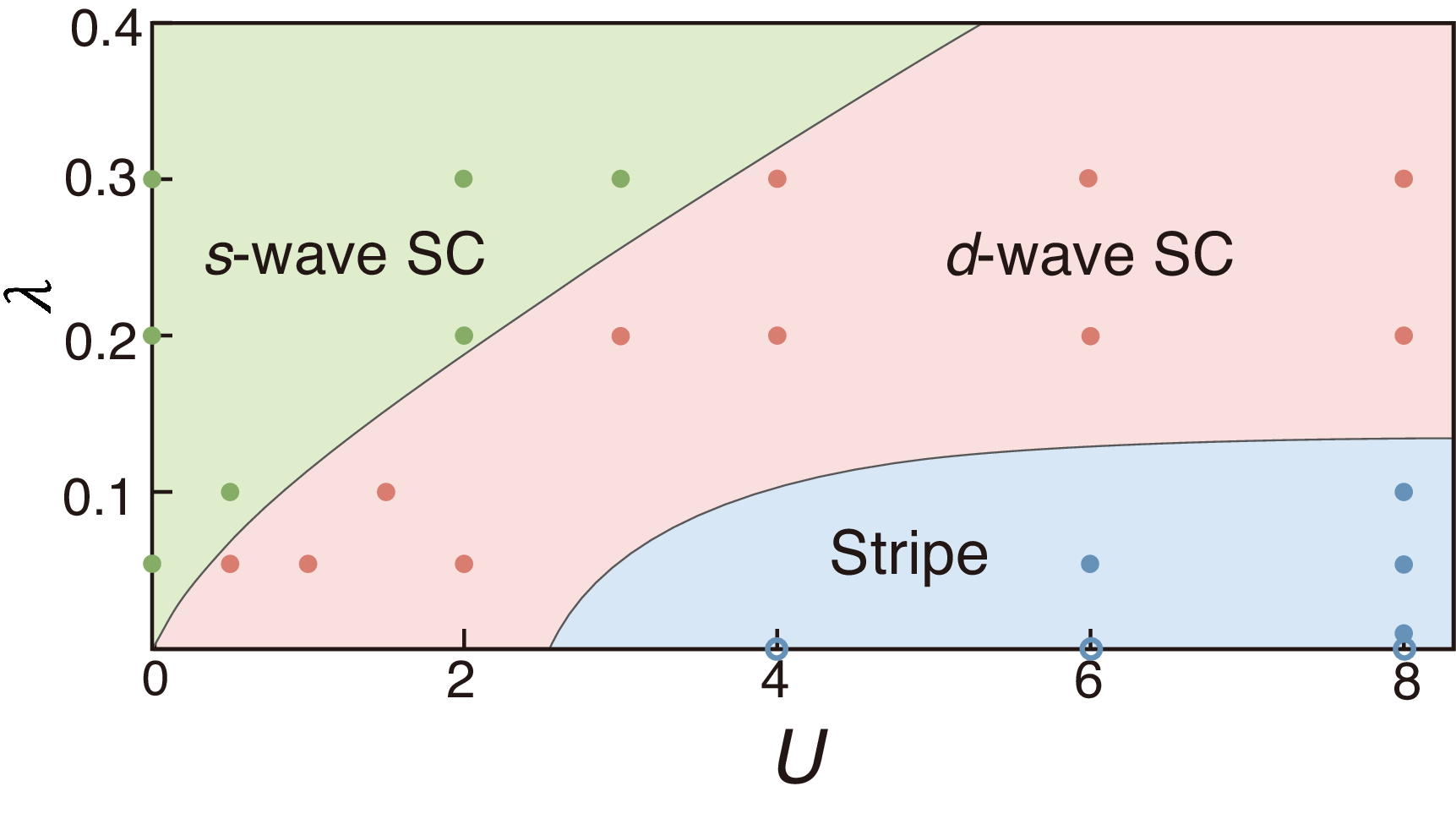}
    \caption{The quantum phase diagram of the Su-Schrieffer-Heeger-Hubbard (SSHH) model as a function of SSH electron-phonon coupling $\lambda$ and Hubbard $U$, obtained from our state-of-the-art DMRG studies. Here, the phonon frequency $\omega_D = 5$ and hole doping concentration $\delta = 1/8$. Robust $d$-wave superconductivity (SC) is obtained by the interplay between strong $U$ and moderate $\lambda$. The solid circles in this figure are DMRG results obtained in the present paper, while the open circles represent results without electron-phonon coupling ($\lambda=0$) obtained in previous works \cite{SimonsCollaboration2020, White2017HybridDMRG}. }
    \label{fig1:phasediagram}
\end{figure}

In this paper, we convincingly show that robust $d$-wave SC can be induced by doping the Su-Schrieffer-Heeger-Hubbard (SSHH) model where the interplay between the Su-Schrieffer-Heeger EPC and the Hubbard interaction is a key for the emergence of $d$-wave SC.
Specifically, we study the interplay between Su-Schrieffer-Heeger (SSH) EPC and the Hubbard interaction in the SSHH model on the square lattice at finite doping by large-scale density-matrix-renormalization-group (DMRG) calculations \cite{White1998, Schollwock2015}.
From state-of-the-art DMRG simulations, we obtained the quantum phase diagram by varying the dimensionless EPC constant $\lambda$ and the Hubbard interaction $U$, as shown in \Fig{fig1:phasediagram}.
Three distinct phases are demonstrated in the quantum phase diagram: $s$-wave SC, $d$-wave SC, and stripe charge-density-wave (CDW) ordering. There is a large region of $d$-wave SC in the phase diagram, indicating the robustness of $d$-wave SC induced by the interplay between SSH and Hubbard interaction. For the $d$-wave SC obtained in the SSHH model, the Luttinger parameter $K_{sc}\approx 0.94$, which implies a divergent SC susceptibility $\chi_{sc}\sim T^{K_{sc}-2}$ as the temperature approaches zero.
In particular, when $U=8t$, the ground state exhibits stripe order for $\lambda<\lambda_c$ but $d$-wave SC for $\lambda>\lambda_c$ with $\lambda_c=0.1\sim 0.2$, which might shed new light to understanding high-temperature SC in cuprates.

{\bf Model:} We consider the SSHH model on the square lattice with the following Hamiltonian
\begin{eqnarray}
&&H=-t \sum_{\langle i j\rangle}\left(c_{i \sigma}^{\dagger} c_{j \sigma}+h.c.\right)
+U \sum_{i} n_{i \uparrow} n_{i \downarrow} \nonumber\\
 && ~+\sum_{\langle i j\rangle} \left( \frac{P_{i j}^{2}}{2 M}+\frac{K}{2} X_{i j}^{2} \right) +\alpha\sum_{\langle ij\rangle} X_{ij}\left(c_{i \sigma}^{\dagger} c_{j \sigma}+h.c.\right),~~
\end{eqnarray}
where $\langle ij\rangle$ denotes nearest-neighbor (NN) bond, $c_{i\sigma}^\dagger$ creates an electron on site $i$ with polarization $\sigma$ = $\uparrow$/$\downarrow$, $n_{i\sigma}=c^\dag_{i\sigma}c_{i\sigma}$, $X_{ij}$ and $P_{ij}$ are the displacement and momentum operators of the optical SSH phonon on the NN bond $\langle ij\rangle$.
The SSH phonon frequency $\omega_D = \sqrt{K/M}$ and the strength of EPC is defined as $\lambda = \alpha^2/(KW)$, where $W=8t$ is the typical band width of the square lattice.
In the following, we set $t=1$ as the energy unit and focus on repulsive Hubbard interaction $U>0$ with hole doping concentration $\delta = 1/8$ and phonon frequency $\omega_D = 5$. 
We believe the results obtained in the present work apply more generally.

To gain some insights, we first discuss the SSHH model in the anti-adiabatic (AA) limit ($\omega_D \to \infty$), although our study below will only focus on finite $\omega_D$ which is more related to real materials. In the AA limit, the SSHH model can be reduced to the following effective model by integrating out phonons
\begin{eqnarray}
&& H_{\mathrm{AA}} = -t \sum_{\langle i j\rangle}\left(c_{i \sigma}^{\dagger} c_{j \sigma}+\mathrm{H}.c.\right)
    +U \sum_{i} n_{i \uparrow} n_{i \downarrow} \nonumber \\
&&~~+J \sum_{\langle i j\rangle}\Bigg[ \mathbf{S}_{i} \cdot \mathbf{S}_{j} +\frac{1}{4} n_i n_j -\frac{1}{2} \left(\Delta_i^\dagger \Delta_j +h.c.\right)  \Bigg],
\end{eqnarray}
where $J=2\alpha^2/K$ is the strength of interactions mediated by optical phonons in the AA limit when $U$ is not considered, $\mathbf{S}_{i}$ is the spin-1/2 operator,
$\Delta_i^\dagger=c^\dagger_{i\uparrow}c^\dagger_{i\downarrow}$ is the on-site pair creation operator, and
$n_i=n_{i\uparrow}+n_{i\downarrow}$ is the density operator.
For finite doping, it is clear that the pair hopping term $-J/2(\Delta^\dag_i\Delta_j+h.c.)$ favors onsite $s$-wave pairing.
However, such onsite $s$-wave pairing will be strongly suppressed by the Hubbard repulsion $U$. Fortunately, the spin exchange coupling $J\mathbf{S}_i\cdot\mathbf{S}_j$ can favor $d$-wave pairing on bonds which cannot be suppressed by the onsite Hubbard repulsion $U$, yielding the possibility of a $d$-wave SC when the $s$-wave components are sufficiently suppressed by the onsite repulsive $U$. The anti-adiabatic limit model here is quite different from previous related studies; bond phonons are considered here while site phonons were used previously \cite{Assaad1996,Assaad1997}. 

For the general SSHH model with finite $\omega_D$ and finite $U$, we employ DMRG 
to study the ground state properties of the model on a $L_x \times L_y$ lattice with open boundary condition (OBC) along the $x$-direction but periodic boundary conditions (PBC) along the $y$-direction, where $L_x$ and $L_y$ denote the lattice size along $x$ and $y$ directions, respectively. Throughout this study, we restrict \(L_y \leq 4\) due to constraints of computational complexity, and all data presented in the main text are based on \(L_y = 4\).
SC properties of the model under study are diagnosed by computing equal-time pair-pair correlation functions in the ground state:
\begin{equation}
    \Phi_{\alpha\beta}(r) =\frac{1}{L_y} \sum_{y=1}^{L_y} \langle \Delta_\alpha^\dagger(x_0,y)\Delta_\beta(x_0+r,y)\rangle,
\end{equation}
where $\Delta^\dagger_{\alpha}(x,y) = \frac{1}{\sqrt{2}}[c^\dagger_{(x,y),\uparrow} c^\dagger_{(x,y)+\alpha,\downarrow} - c^\dagger_{(x,y),\downarrow} c^\dagger_{(x,y)+\alpha,\uparrow}] $ is the spin-singlet pair creation operator on site ($\alpha=0$) or on NN bond ($\alpha = \hat x, \hat y$) and $x_0$ is the reference point.
To minimize the boundary effect of the finite system, for various $L_x$, we calculate the $\Phi_{\alpha\beta}(r=\frac{L_x}{2}-1)$ with $x_0$ around $L_x/4$. Density-density and spin-spin correlation functions are computed similarly.
In our calculations, we keep the bond dimensions up to $D=18000$ to obtain reliable results with typical truncation error $\varepsilon \sim 2 \times 10^{-6}$.
All the data shown in the plots have been extrapolated to the $\varepsilon \rightarrow 0$ limit using quadratic fitting. Numerical details are discussed in the Supplemental Materials \cite{Supplementary}.

The main results of this work are illustrated in the quantum phase diagram of SSHH model as a function of Hubbard $U$ and the dimensionless SSH EPC constant $\lambda$, as shown in Fig.~\ref{fig1:phasediagram}. We identify three phases according to their long-range charge and pairing properties: (a) an on-site {\it s}-wave SC phase characterized by dominant quasi-long-range SC correlation and subdominant quasi-long-range charge density correlation; (b) The {\it d}-wave SC phase characterized by dominant SC correlation with {\it d}-wave symmetry and subdominant charge-density correlation; (c) stripe CDW phase with short-range SC correlation and long-range CDW ordering.

{\bf The $s$-wave SC for finite $\lambda$ and zero $U$:} When $U=0$, there is only SSH electron-phonon interaction, and the SSHH model is reduced to the SSH phonon model.
At half filling, the ground state of the SSH phonon model is either AFM or valence bond solid (VBS) depending on the value of $\lambda$. It is expected that doping holes into the SSH phonon model can induce SC.
We first performed our DMRG calculations of the SSH phonon model (namely $U=0$) at finite doping ($\delta=1/8$) for various values of EPC coupling $\lambda$.

\begin{figure}
    \centering
    \includegraphics[width = 0.45\textwidth]{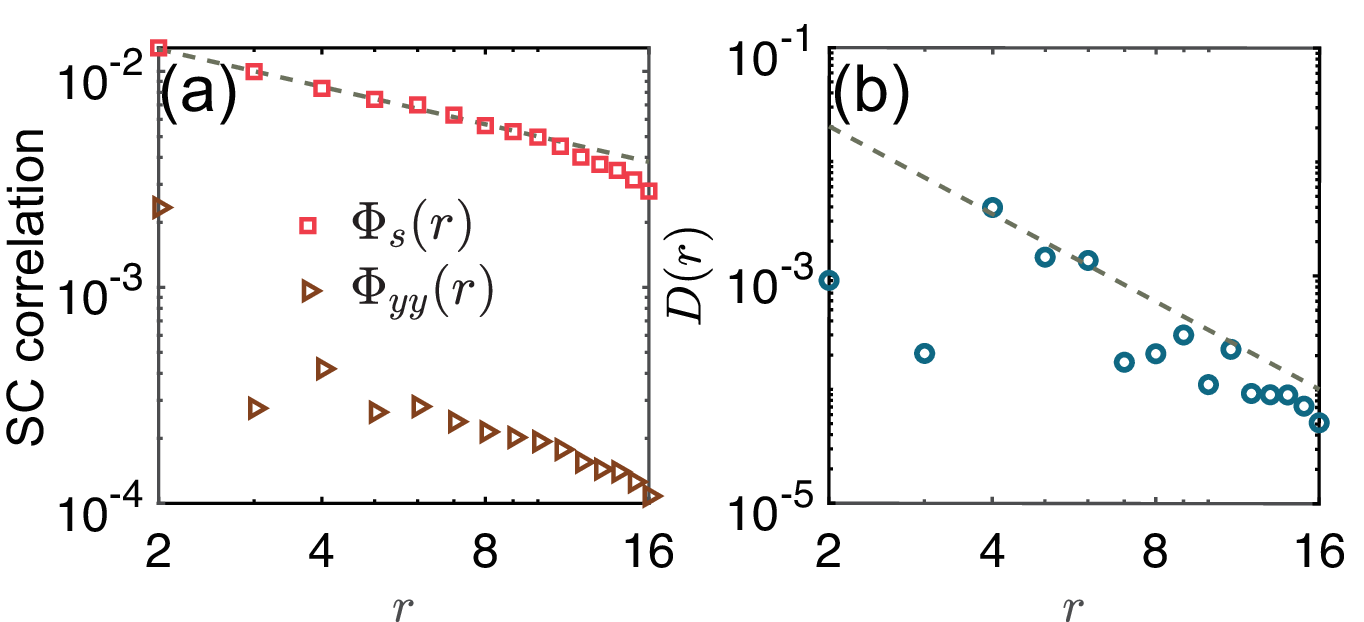}
    \caption{Properties of the $s$-wave SC phase for the SSHH model with $\lambda=0.3$ and $U=0$ on $L_x=48$ cylinder. (a) Onsite pair-pair and bond pair-pair correlations are plotted on a double-logarithmic scale. (b) Density-density correlation is plotted using a double-logarithmic scale.}
    \label{fig2:physicalpropertiesins-wave}
\end{figure}

Our DMRG calculations of the model with $U=0$ and various $\lambda$ showed that the onsite $s$-wave pairing is always the most dominant pairing channel, as shown in Fig.~\ref{fig1:phasediagram}. Besides the dominant onsite $s$-wave pairing $\Phi_s$, there is also subdominant bond $d$-wave pairing $\Phi_{yy}$, as shown in Fig.~\ref{fig2:physicalpropertiesins-wave}(a), which can be understood heuristically from the AA limit.
In the AA limit, the effective pairing hopping term $-J/2(\Delta^\dag_i\Delta_j+h.c.)$ which favors onsite $s$-wave pairing is more effective in inducing SC than the effective AF spin-exchange interaction $J\mathbf{S}_i\cdot\mathbf{S}_j$ which tends to favor bond $d$-wave pairing.
For $\lambda=0.3$ and $U=0$, we find that the leading SC correlation function is the quasi-long-range on-site correlation $\Phi_s(r) \propto r^{-K_{sc}}$ with Luttinger exponent $K_{sc} = 0.56 \pm 0.04 $, while the on-bond $\Phi_{yy}(r)$ is roughly two orders of magnitude weaker than $\Phi_{s}(r)$, as as shown in Fig.~\ref{fig2:physicalpropertiesins-wave}.
We also measured the density-density correlation function which is also quasi-long-range, $D(r) \propto r^{-K_{c}}$, with a much larger exponent $K_c = 2.4 \pm 0.2$, as shown in Fig.~\ref{fig2:physicalpropertiesins-wave}(b). The spin and single-particle excitations are gapped in the $s$-wave SC phase (see the SM \cite{Supplementary} for details).

\begin{figure}
    \centering
    \includegraphics[width=0.49\textwidth]{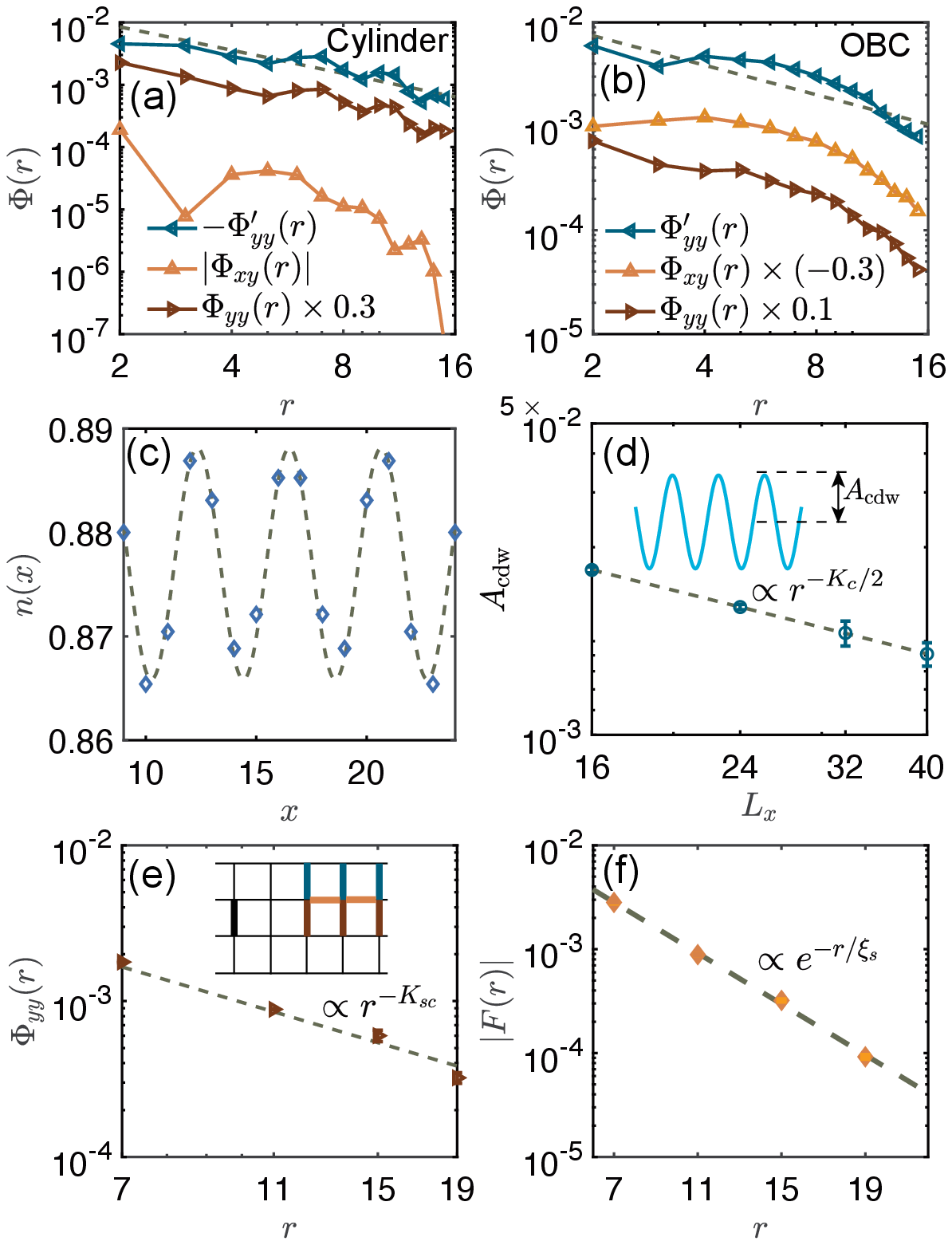}
    \caption{Properties of the $d$-wave SC phase for the SSHH model with $\lambda = 0.3$ and $U=8$, deep inside the $d$-wave SC phase.
(a) The pair-pair correlation functions between different types of bonds $\Phi_{yy}$, $\Phi_{yy}'$, and $\Phi_{yx}$ on $L_x=32$ cylinders. The data for $\Phi_{yy}$ are scaled to lift the curves for clarity.
The grey dashed line shows the fit of $\Phi_{yy}(r)$ with $K_{sc}=0.94 \pm 0.2$.
(b) The same correlation functions as in (a), but for strip geometry, where the fit gives $K_{sc} = 0.77 \pm 0.2$. 
(c) The charge density profile $n(x)$ on the $L_x=32$ cylinder fitted by $n(x) = A_{\mathrm{cdw}} \cos(Q x+\theta)+n_0$ (Grey dashed line). The green circle marks the reference point $x_0$ in the correlation functions.
(d) Finite-size scaling of the CDW amplitude $A_{\mathrm{cdw}}(L_x)$, obtained on cylinders with lengths $L_x=16 \sim 40$. Error bars represent confidence intervals from finite bond dimension extrapolations.
(e) The pair-pair correlation $\Phi_{yy}(r)$ on the $y$-bond as a function of distance $r = L_x/2 - 1$. The inset illustrates the correlation coordinates for \(\Phi_{yy}(r)\), \(\Phi_{yy}'(r)\), and \(\Phi_{xy}(r)\). The bold black bond represents the reference bond, while the colored bonds correspond to the secondary bonds, with colors matching the curves in (a) and (b).
(f) The spin-spin correlation $F(r)$ as a function of distance $r$, shown in a semi-logarithmic plot. Panels (a), (b), and (e) are shown in double-logarithmic scale.}
    \label{figure:d-wavestate}
\end{figure}

{\bf The $d$-wave SC for moderate EPC and large $U$:} When fixing $\lambda$, i.e. $\lambda=0.3$, and gradually increasing $U$ from zero, the onsite Hubbard repulsion naturally suppresses the onsite $s$-wave pairing tendency but leaves the bond $d$-wave pairing channel nearly unaffected. Consequently, we expect that with increasing $U$ the onsite $s$-wave pairing should become increasingly weaker such that the system could be driven into a $d$-wave SC phase by sufficiently large $U$. Note that it was shown that $d$-wave SC with exponentially-small $T_c$ can appear for weak $U$ and zero $\lambda$ from perturbative RG analysis \cite{Kivelson2010RG} as well as for not strong $U$ and $\lambda$ \cite{QHWang-arXiv2022}.  In this paper we focus more on strong $U$ which is the case of cuprates and explore how robust $d$-wave SC can emerge from the interplay between strong $U$ and moderate $\lambda$. Moreover, the SSHH model studied by DMRG in this paper in obtaining $d$-wave SC features finite phonon frequency and finite doping, qualitatively different from the ones previously studied by QMC where infinite phonon frequency and zero doping were considered \cite{Assaad1996,Assaad1997}.

In our DMRG calculation, the competition between $d$-wave and $s$-wave SC is studied by comparing the on-site $s$-wave pair-pair correlation function $\Phi_s(r)$ and the bond pair-pair correlation $\Phi_{yy}(r)$.
For $\lambda=0.3$ and $U=0$, the on-site $s$-wave pair-pair correlation $\Phi_s(r)$ is roughly two orders of magnitude stronger than the bond pair-pair correlation $\Phi_{yy}(r)$.
When $U$ is increased from zero, the amplitude of $\Phi_s(r)$ is quickly suppressed and eventually becomes a sub-dominant correlation than bond $d$-wave pairing when $U\gtrsim 3$ for $\lambda = 0.3$, as shown in the Fig.~\ref{fig1:phasediagram} as well as in the SM \cite{Supplementary}.

We choose $\lambda = 0.3$ and $U = 8$, which is deep inside the $d$-wave SC phase, as a representative model to investigate the properties of this $d$-wave SC phase and its possible relevance to cuprates.
For dominant bond pair-pair correlations, aside from the extended $s$-wave and usual $d$-wave pairing the special geometry of the 4-leg cylinder also allows the appearance of ``plaquette'' $d$-wave pairing, as pointed out in Refs. \cite{Dodaro2017, White2020PlaquetteDwave}.
To distinguish between these three candidate pairing  symmetries, we further measure the pair-pair correlation functions $\Phi_{yy}^{\prime}(r)$ on cylinder and on strip geometry with open boundary conditions along both directions, respectively.
Here $\Phi_{yy}^{\prime}$ is pair-pair correlations on two bonds which are separated by distance $r$ along the x direction and distance one along y direction. In the inset of Fig.~\ref{figure:d-wavestate}(e), we show the correlation coordinates of $\Phi_{yy}^{\prime}(r)$ as the black and green bonds.
In Fig.~\ref{figure:d-wavestate}(a) and Fig.~\ref{figure:d-wavestate}(b), we plot $\Phi_{yy}$, $\Phi_{yy}^{\prime}$, and $\Phi_{yx}$ measured on the $L_x=32$ cylinder and strip, respectively.
On both systems, we observe the quasi-long-range SC correlation at long distance with exponent $K_{\text sc}\approx 0.94$ for cylinder and $K_{\text sc}\approx 0.77$ for strip, respectively.
However, their pair symmetries are different.
On the 4-leg cylinder, the fact that $\Phi_{yy}(r) \simeq -\Phi_{yy}^{\prime}(r) \gg |\Phi_{yx}(r)| > 0$ implies that the system forms a plaquette $d$-wave SC state which shares similar properties with the one reported in the pure Hubbard model with negative next-nearest-neighbor $t'$ term.
While on the 4-leg strip, it is clearly ordinary $d$-wave pairing with $\Phi_{yy}(r) \simeq \Phi_{yy}^{\prime}(r) \simeq -\Phi_{yx}(r)$, which implies that the $d$-wave pairing induced by the interplay between EPC and Hubbard interaction is the usual $d$-wave on wider and 2D systems where the 4-site special plaquette relevant only to 4-leg ladders no longer exists.

We further study the charge-density properties of the representative model with $U=8$ and $\lambda=0.3$ by measuring the density profile $n(x) = \frac{1}{L_y} \sum_{y=1}^{L_y} \langle n(x,y)\rangle $ in the bulk of the systems.
As shown in Fig.~\ref{figure:d-wavestate}(c), the density profile exhibits a clear ``half-filled'' stripe with wave-length 4 for doping $\delta=1/8$, i.e., half a doped hole in each CDW unit cell.
The low energy physics in the charge sector can be determined by the long distant behavior of density-density correlation.
For the ground state with gapless charge excitations, the spatial decay of the correlation is dominated by a power-law function with charge exponent $K_c$.
On the finite system, the Luttinger parameter $K_c$ can be extracted from $A_{\mathrm{cdw}}(L_x) \propto L_x^{-K_c/2}$, where $A_{\mathrm{cdw}}(L_x)$ is the CDW amplitude in the middle of the cylinders with length $L_x$.
As illustrated in Fig.~\ref{figure:d-wavestate}(d), the $A_{\mathrm{cdw}}$ obtained on a series of cylinders with $L_x=16 \sim 40$ show a clear power-law decaying with exponent $K_c=1.35 \pm 0.01$, indicating the charge gap vanishes in the $d$-wave phase.
We also observe consistent results from the Friedel oscillation induced by the open boundary of the cylinder ( details are provided in the SM \cite{Supplementary}). We further investigated the magnetic properties of the ground state by measuring the spin-spin correlation functions defined as $F(r) =\frac{1}{L_y} \sum_{y=1}^{L_y}\Vec{S}_{x_0,y} \cdot \vec{ S}_{x_0+r,y}$, where $\Vec{S}_{x,y}$ is the spin operator on site $(x,y)$. Following the same procedure used for SC correlation, we find a short-range spin-spin correlation $F(r) \propto \exp(-r/\xi_{s})$ with correlation length $\xi_{s} = 3.54$ as shown in Fig.~\ref{figure:d-wavestate}(f), indicating a finite spin gap of the $d$-wave SC phase at $U=8$ and $\lambda=0.3$.

\begin{figure}
    \centering
    \includegraphics[width = 0.45\textwidth]{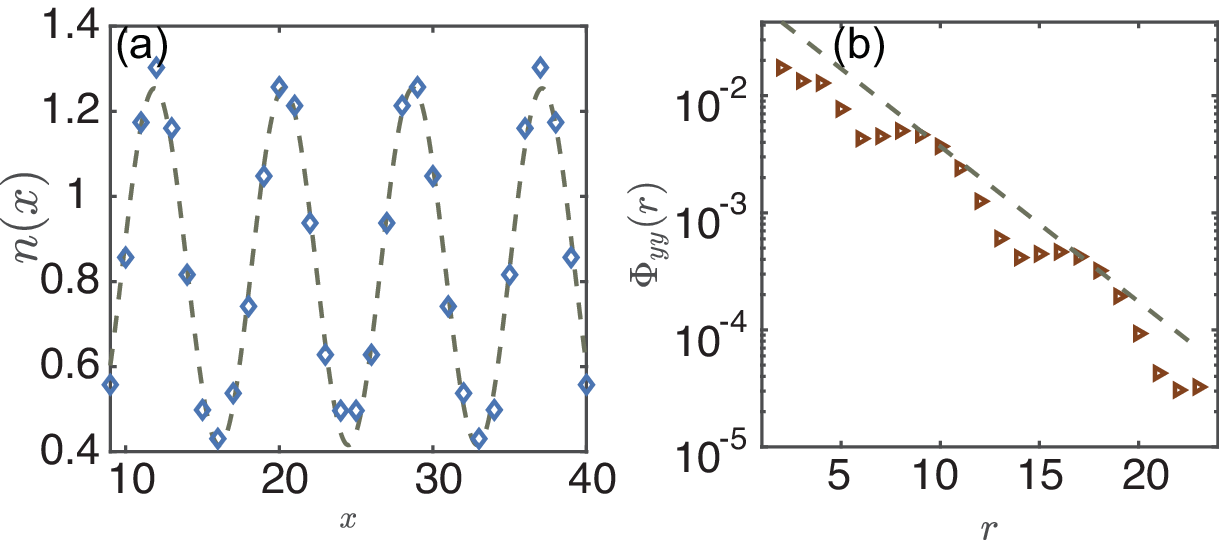}
    \caption{Properties of the stripe phase for strong $U$ and small $\lambda$. (a) Charge density profile of $\lambda=0.1$ and $U=8$ model on $L_x = 48$ cylinder. The dominant CDW order with wavelength $8$ is formed at $\delta=1/8$ doping. (b) SC pair-pair correlation function in the semi-logarithmic plot for the same system.}
    \label{fig:CDWstate}
\end{figure}

{\bf The stripe phase for strong $U$ and $\lambda<\lambda_c$:} There were plenty pieces of evidence that the ground state of the pure Hubbard model (namely $\lambda=0$) with doping $\delta=1/8$ and moderate $U$ ($U= 4\sim 8$) exhibits a long-range filled-stripe CDW order \cite{SimonsCollaboration2015, SimonsCollaboration2017, SimonsCollaboration2020, White2017HybridDMRG}.
Our DMRG study also found this filled-stripe state characterized by the long-range CDW order with ordering momentum $Q_{\mathrm{CDW}}=2\pi/8$ and short-range SC correlation on four-leg cylinders for moderate $U$.
We then study the fate of this filled-stripe state by increasing EPC $\lambda$ while fixing $U$.
For instance, when $U=8$ we find that the filled-stripe phase is stable against small $\lambda$, e.g., the SC correlation function $\Phi_{yy}(r) \propto e ^{-r/\xi_{sc}}$ still decays exponentially with a long correlation length $\xi_{sc}=3.3$ for $\lambda=0.1$, as shown in Fig.~\ref{fig:CDWstate}.
When EPC is further increased to $\lambda \ge 0.2$, the system is driven into the $d$-wave SC phase by the SSH phonon.
Consequently, for $U=8$ and $\lambda>\lambda_c$ with $\lambda_c=0.1\sim 0.2$, robust $d$-wave SC emerges.
This strong coupling $d$-wave SC phase is the key result of this paper, which implies that the interplay between Hubbard repulsion and SSH electron-phonon coupling plays a crucial role in inducing $d$-wave SC.

{\bf Discussion and concluding remarks:}
In conclusion, from the state-of-the-art DMRG calculations, we have shown convincing evidence that the interplay between moderate SSH-type EPC and strong Hubbard $U$ on the square lattice can induce robust $d$-wave SC, which cannot be obtained from either moderate EPC or strong Hubbard repulsion alone. The onsite $s$-wave SC is induced when there is only EPC, while the stripe CDW order is obtained for strong Hubbard interaction without EPC.
We also established the dominant SC correlations and subdominant CDW correlations by finite-size scaling in the $d$-wave SC phase, with $K_{sc}\approx 0.94$ and divergent SC susceptibility.
Although the error bar in numerical values of the Luttinger exponents obtained from DMRG is not that small, they are already close to the physical properties of a Luther-Emery liquid, and the $d$-wave SC state we obtained exhibits most of the qualitative properties of a Luther-Emery liquid \cite{LutherEmery1974, LeonPRB1996WeakCoupling, Kivelson2004Mechanism, Gannot2020Hubbardladder}.
To determine those exponents more accurately, it is desired to study the model on ladders with much larger $L_x$ and keep more states in DMRG, which can be explored in the future.

The $d$-wave SC states obtained for strong $U$ and moderate $\lambda$ could have potential relevance for understanding the $d$-wave SC in cuprates which have both strong onsite Coulomb repulsion and moderate EPC.
For $U=8t$, our DMRG calculations showed that $d$-wave SC phase emerges for $\lambda>\lambda_c$ with $\lambda_c=0.1\!\sim\! 0.2$. For cuprates, recent experiments estimate that its EPC is about $\lambda = 0.2\!\sim\!0.3$ (see, e.g. Refs. \cite{ZXShen2018, ZXShen2021_1Dcuprate, YaoWang2021}). 
Moreover, we think that our findings presented in this paper might pave a new route in looking for unconventional SC and high-temperature SC in quantum materials with corroborate electron-phonon and electron-electron interactions.

{\it Acknowledgement}: We would like to thank Xun Cai, Steven Kivelson, Zi-Xiang Li, Yi-Ming Wu, Zhengzhi Wu, and Rong-Yang Sun for helpful discussions. This work is supported in part by the NSFC under Grant No. 11825404 (HXW and HY), the MOSTC Grant No. 2021YFA1400100 (HY), Shanghai Pujiang Program under Grant No.21PJ1410300 (YFJ). The DMRG code used to simulate the SSHH model in this work is publicly available in Github \cite{code}.

%

 \widetext
 \section{Supplemental Materials}

 \setcounter{equation}{0}
 \setcounter{figure}{0}
 \setcounter{table}{0}
\makeatletter
 \renewcommand{\theequation}{S\arabic{equation}}
 \renewcommand{\thefigure}{S\arabic{figure}}
\renewcommand{\bibnumfmt}[1]{[S#1]}
\renewcommand{\citenumfont}[1]{S#1}

\subsection{A. Numerical details of the DMRG calculation}
\begin{figure}[hb]
    \centering
    \includegraphics[width=0.3\textwidth]{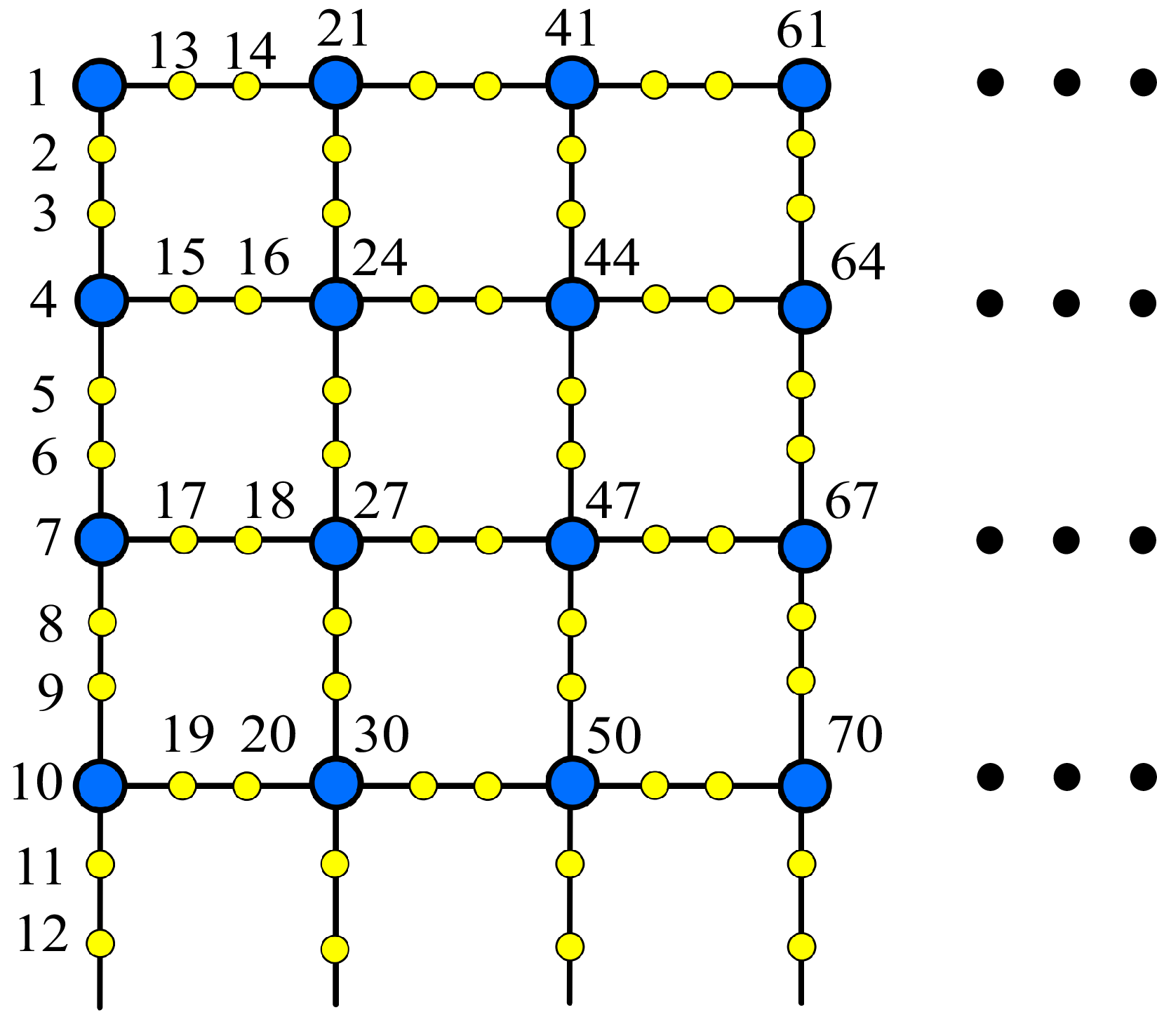}
    \caption{ The lattice geometry of $4 \times 4$ cylinder with $M=2$ pseudo-sites per link. The blue circles represent fermion sites, while the yellow circles represent the pseudo-sites. The mapping of the cylinder to 1D MPS is arranged according to the labels of the sites.}
    \label{fig:MPS order}
\end{figure}

Following the pseudo-site scheme in Ref. [84] of the main text, we cut off the dimension of the local Hilbert space of the SSH phonon on each link to $2^M$ and then decompose the $2^M$ dimensional Hilbert space to $M$ pseudo-sites of hard-core bosons. Fig.~\ref{fig:MPS order} shows the lattice geometry of the SSHH model with two phonon pseudo-sites per link (the yellow circles), and the numbers around the (pseudo-)sites demonstrate how we map a 2D cylinder into a 1D matrix product state (MPS). Only the pseudo-sites with labels smaller than 21 are marked for clarity.

Because all of the electron sites are surrounded by phonon pseudo-sites that do not involve the total electron number $N$ or spin momentum $S_z$ of electrons, the quantum numbers of the MPS' auxiliary bond cannot ascent automatically, even with the two-site update algorithm. To circumvent this issue of the EPC system, we implement the subspace expansion method [85] in the two-site update. By combining these two methods, we are able to obtain more accurate results at finite bond dimension $D$. In our code, we also parallel the contraction of the tensor to access higher bond dimension $D$ and more reliable extrapolation of physical quantities as $D\rightarrow \infty$, or equivalently truncation error $\epsilon \rightarrow 0$.

\subsection{Supplemental data for the $s$-wave state and the $d$-wave state}
\begin{figure}[htbp]
    \centering
    \includegraphics[width=0.3\textwidth]{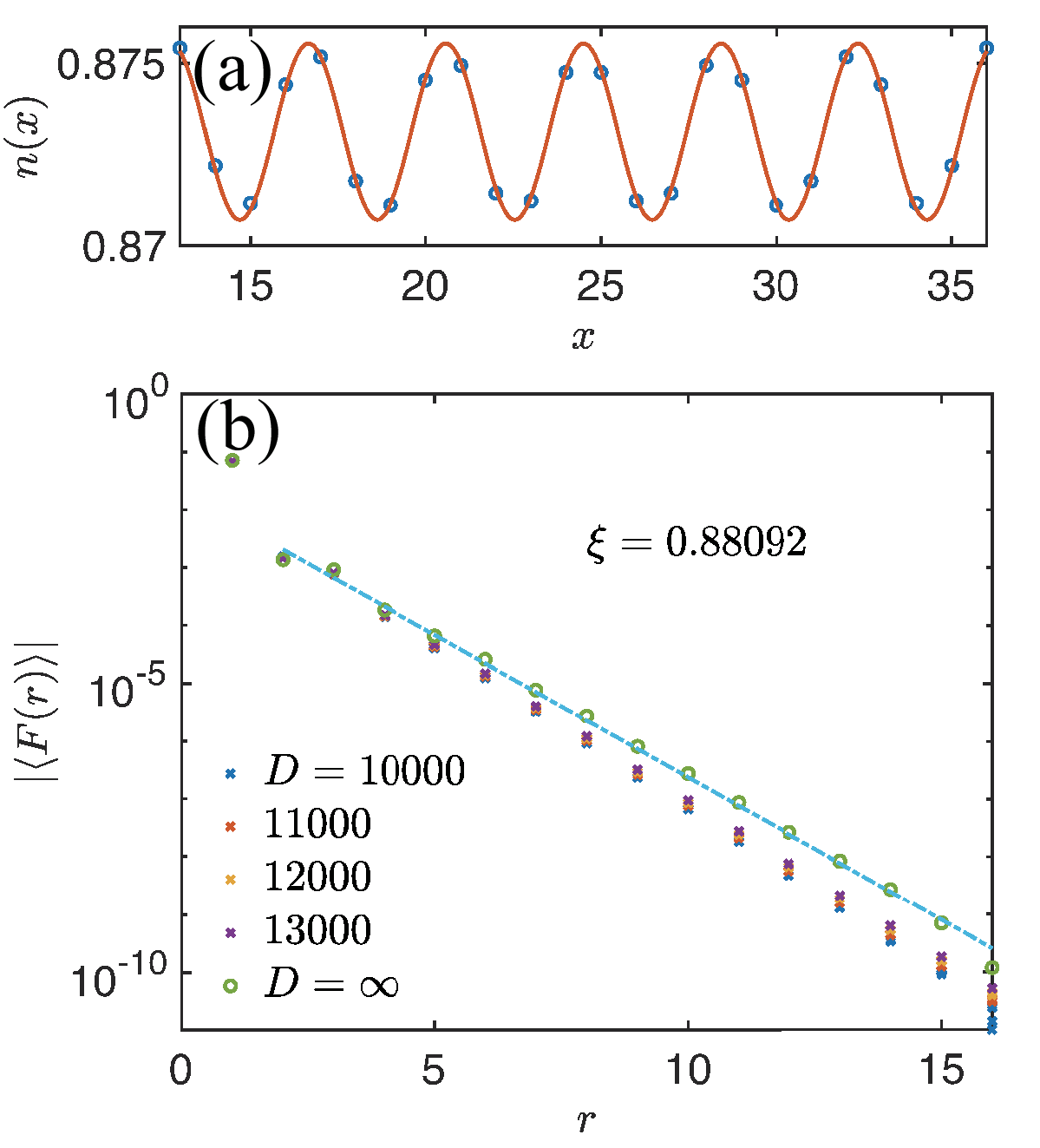}
    \caption{(a) The charge density profile and (b) spin-spin correlation function of $s$-wave SC state on $L_x = 48$ cylinder with model parameter $U=0$ and $\lambda=0.3$.}
    \label{fig:SM s-wave}
\end{figure}

In this part, we show more results of observation as a complement to the main text. Fig.~\ref{fig:SM s-wave} show the charge density profile and spin-spin correlation function of the $s$-wave state with model parameters $\lambda=0.3$ and $U=0$. Like the $d$-wave SC state, the period of the charge density distribution is 4 lattice constants. The spin correlation function decays exponentially with correlation length $\xi_s = 0.87$. Such a short correlation length indicates a spin gap in the $s$-wave SC ground state, consistent with the Luther-Emery liquid discussed in the main text.
\begin{figure}
    \centering
\includegraphics[width = 0.4\textwidth]{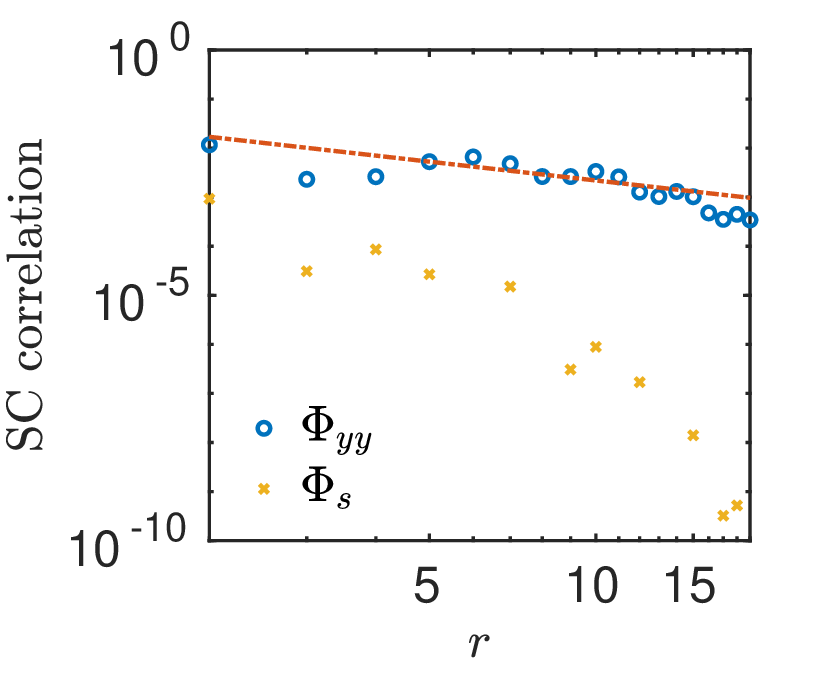}
\includegraphics[width = 0.45\textwidth]{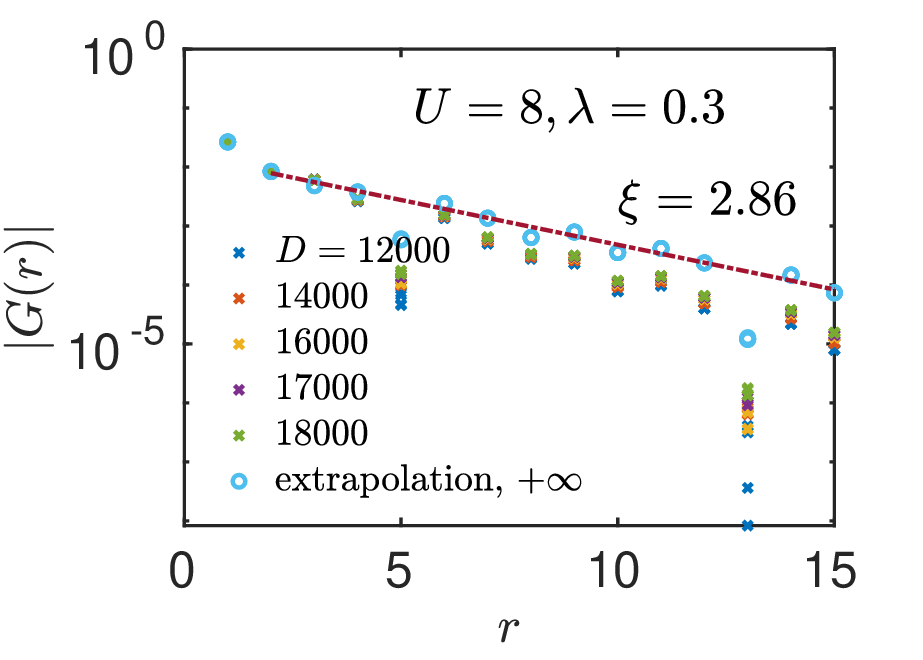}
    \caption{Left panel: SC correlation $\Phi_{yy}(r)$ and $\Phi_{s}(r)$ of $L_x=40$ cylinder in \dwave state. The red dashed line is fitting of the exponent $K_{sc} = 1.27$. Right panel: Single particle correlation $G(r)$ of $L_x = 32$ cylinder in \dwave state.}
    \label{fig: d-wave state Lx40}
\end{figure}

In Fig.~\ref{fig: d-wave state Lx40}, we show the SC correlations and single particle correlation for the $d$-wave state at point $\lambda = 0.3$ and $U=8$ in the phase diagram. The left panel shows $\Phi_{yy}$ and $\Phi_{s}$ in the $L_x=40$ cylinder. The $d$-wave SC correlation $\Phi_{yy}$ decays according to a power law with the exponent $K_{sc} \simeq 1.2$. The exponent is slightly larger than that in the main text since the calculation does not completely converge with respect to the bond dimensions. We also discover an exponentially decaying $s$-wave SC correlation $\Phi_{s}$, attributed to the suppression effect caused by the large on-site Coulomb repulsion $U$. The right panel shows the single particle correlation function $G(r) = \langle c^\dagger(x_0, y) c(x_0 + r, y) \rangle$ decaying exponentially fast with correlation length $\xi \approx 2.86$. It indicates a finite superconductivity gap.

\subsection{Evolution of pairing symmetry from $s$-wave to $d$-wave SC state}
\begin{figure}[tb]
    \centering
    \includegraphics[width=\textwidth]{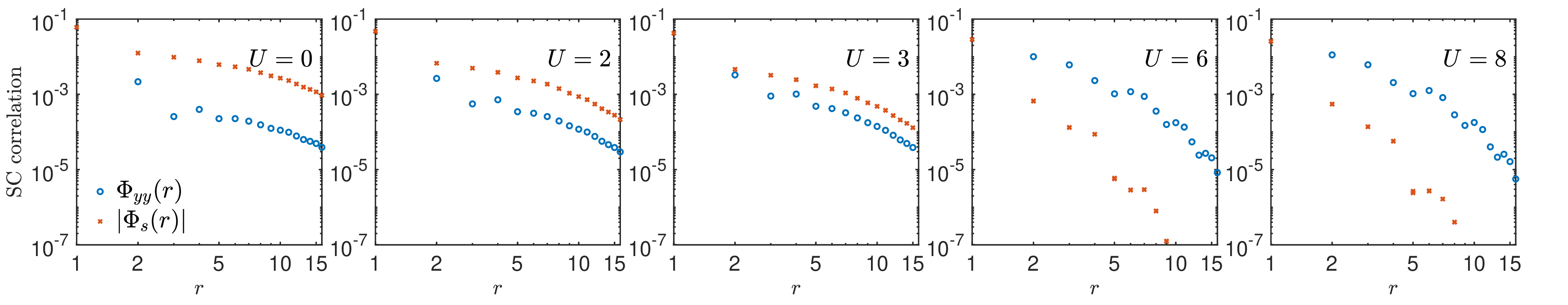}
    \caption{The competitions between the $s$-wave pairing and the $d$-wave pairing when Hubbard repulsion $U$ increases from $0$ to $8$ with fixed $\lambda = 0.3$. The data are obtained with finite bond dimensions $D = 8000 \sim 10000$. When $U$ increases, the $s$-wave SC correlation is significantly suppressed, while the $d$-wave SC correlation remains nearly unchanged.}
    \label{fig:pairing evolution}
\end{figure}

The Hubbard repulsion $U$, as mentioned in the main text, can significantly suppress the $s$-wave SC component while leaving the $d$-wave SC almost unchanged. Thus, a sufficiently large $U$ can drive the system from $s$-wave SC state to $d$-wave SC state. To depict the evolution of the pairing symmetry as $U$ increases, we calculate the SC correlations under different $U$ in the $s$-wave SC phase and the $d$-wave phase with fixed EPC constant $\lambda = 0.3$. The results are shown in Fig.~\ref{fig:pairing evolution}. For $U \leq 3$, the amplitudes of the $s$-wave SC correlation $\Phi_s(r)$ are stronger than those of the $d$-wave SC correlation $\Phi_{yy}(r)$, and both of them decay in the power law. The difference between the amplitudes of the $s$-wave and $d$-wave SC are decreasing as $U$ approaches to $3$. For large $U\ge6$, the $s$-wave SC correlation decays much faster than $d$-wave SC correlation because of the suppression from the Coulomb repulsion. The evolution of the pairing symmetry supports the mechanism of the $d$-wave SC state discussed in the main text.

\subsection{Finite bond dimension extrapolations}
\begin{figure}[tb]
    \centering
    \includegraphics[width=0.6\textwidth]{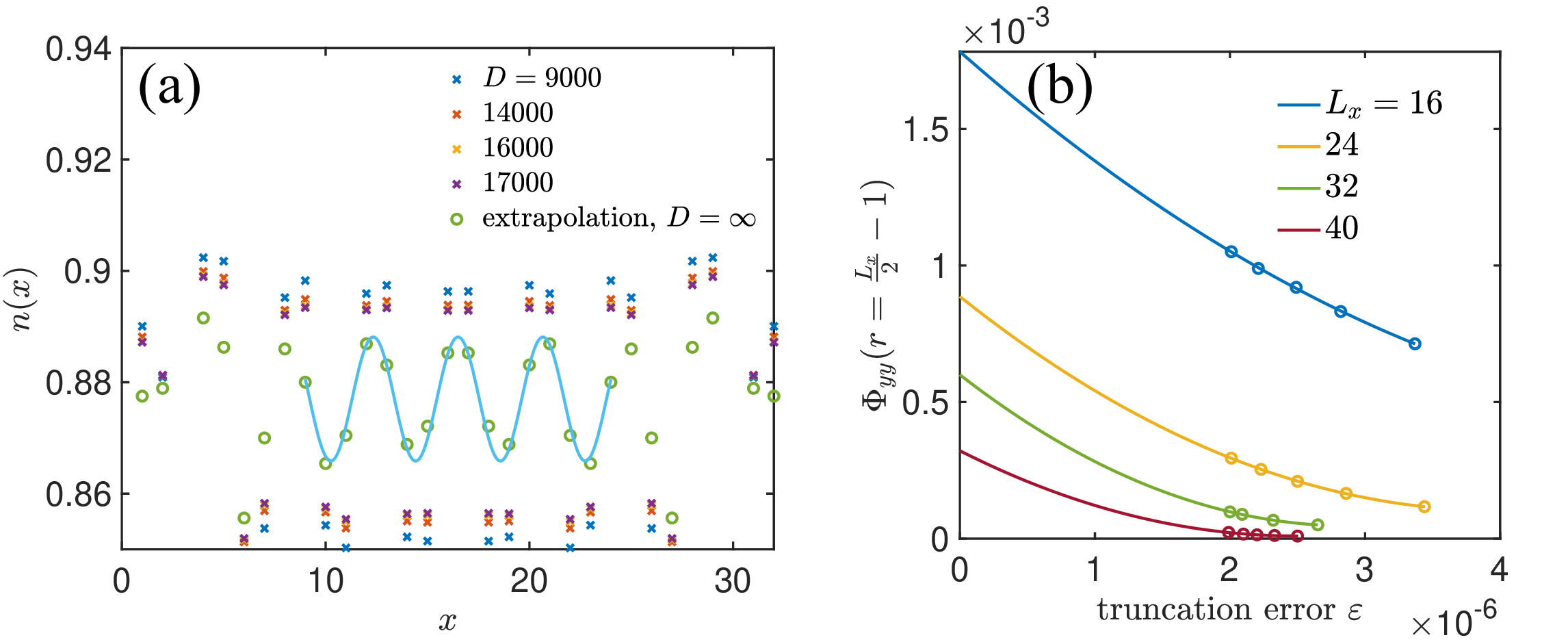}
    \caption{Examples of finite truncation error extrapolation in $d$-wave SC phase with $U=8, \lambda = 0.3$. (a) The charge density profile on $L_x=32$ cylinder, where cross markers represent the data at finite bond dimensions, and circles represent the data after extrapolation. The CDW amplitude $A_{\mathrm{CDW}}$ is extracted from the blue curve in the central part of the cylinder by Eq.~\ref{Eq:SM1}.
    (b) The extrapolation of SC correlation functions $\Phi_{yy}(r = L_x/2 -1)$ on systems with different size.}
    \label{fig: extrapolation in d-wave state}
\end{figure}

In this work, we use the second-order polynomial function $O(\epsilon)=A \epsilon^2+B\epsilon+C$ to extract the physical quantities $O(\epsilon\rightarrow0)$, or equivalently $O(D\rightarrow \infty)$, where $\epsilon$ is the truncation error in the middle of the system for a given bond dimension $D$.
In Fig.~\ref{fig: extrapolation in d-wave state}, we show two examples of the finite truncation error extrapolations of the charge density profile and SC correlation functions in the $d$-wave phase.

In Fig.~\ref{fig: extrapolation in d-wave state}(a), we calculate the electron density $n(x)$ with $D$ varied from $9000$ to $17000$ and extrapolate $n(x, D=\infty)$ for each $x$ using the second order polynomial function. The CDW amplitude $A_{\mathrm CDW}(L_x)$ is extracted from $n(x, D=\infty)$ in the middle of the cylinders by fitting the function
\begin{equation}
    n(x) = A_{\mathrm{CDW}} \cos(Q\cdot x+\theta) + n_0,
    \label{Eq:SM1}
\end{equation}
where $Q$ is the CDW ordering vector, $n_0 = 1-\delta$ average charge density, $\theta$ is the non-universal phase factor. The result of $L_x$ dependence of CDW amplitude $A_{\mathrm CDW}(L_x)$ is shown in Fig. 2(c) in the main text.
The extrapolation of SC correlation function $\Phi_{yy}(r = L_x/2 -1)$ on systems with different $L_x=16\sim 40$ is directly shown in Fig.~\ref{fig: extrapolation in d-wave state}(b). For the largest $L_x=40$ system, a much larger bond dimension is required to obtain reliable extrapolation. Limited by the computational resource, the accuracy of the $\Phi_{yy}$ on $L_x=40$ cylinder is not as good as the one on the shorter cylinders. Consequently, the SC correlation $\Phi_{yy}(L_x/2-1)$ with $L_x=40$ deviates from the power law decay slightly, as shown in Fig.~3(c) in the main text.

\subsection{The Holstein-Hubbard model on square lattice}
The Holstein-Hubbard model is defined as
\begin{equation}
    \begin{aligned} {H}=&-t \sum_{\langle i, j\rangle, \sigma}\left({c}_{i, \sigma}^{\dagger} {c}_{j, \sigma}+\text { H.c. }\right) \\ &+\frac{U}{2} \sum_{i} {n}_{i}^{2}+\alpha \sum_{i} {n}_{i} \hat{X}_{i}+\sum_{i}\left[\frac{\hat{P}_{i}^{2}}{2 m}+\frac{k \hat{X}_{i}^{2}}{2}\right], \end{aligned}
\end{equation}
where the phonon lives on site and couples with on-site electron density $n_i$. In the anti-adiabatic limit, one can integrate the optical phonon and get
\begin{equation}
    H_{\mathrm{AA}} = -t \sum_{\langle i, j\rangle, \sigma}\left({c}_{i, \sigma}^{\dagger} {c}_{j, \sigma}+\text { H.c. }\right) +\frac{U_{eff}}{2} \sum_{i} {n}_{i}^{2}
\end{equation}
with effective interaction $U_{eff} = U - \alpha^2/k$. Here, the instantaneous phonons only renormalize the Hubbard repulsion $U$ to a weaker value.

\begin{figure}[htb]
    \centering \includegraphics[width=0.3\textwidth]{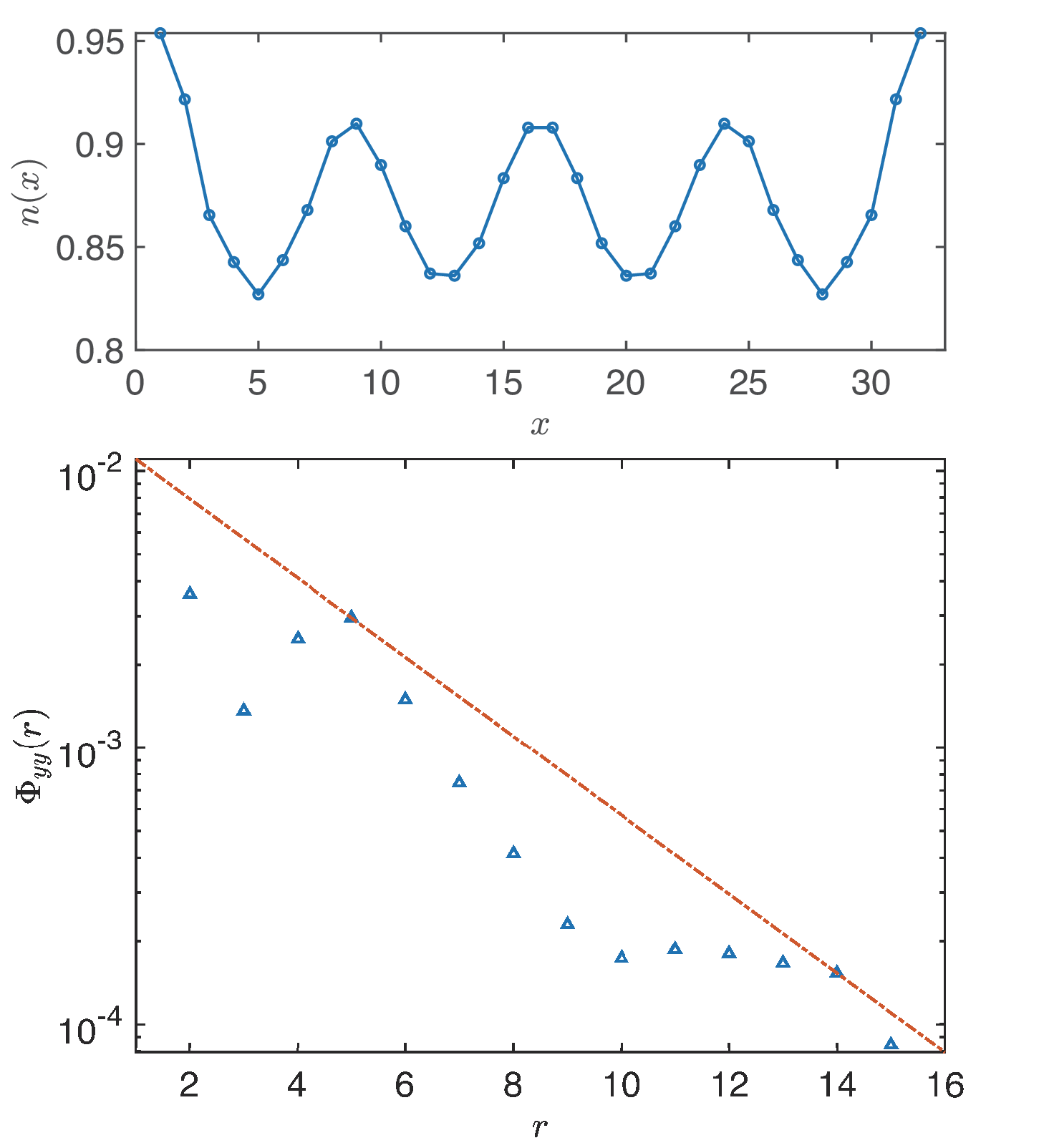}
    \caption{Charge density profile and SC correlation for the Holstein-Hubbard model on $L_x=32$ cylinder with model parameters $U=8$ and $\lambda=0.2$. The data are obtained at finite keep state $D=14000$. }
    \label{SM_fig: Holstein CDW}
\end{figure}

We calculate the Holstein-Hubbard model with $\omega = 5$, $\lambda=0.2$, and $U=8$ on four-leg cylinders. At doping level $\delta = 1/8$, the ground state is a filled-striped state, the same as the one of the pure Hubbard model. The charge density profile and the SC correlation functions of the ground state are shown in Fig.~\ref{SM_fig: Holstein CDW}. The wavelength of CDW order is 8 lattice constants and the SC correlation decays exponentially fast correlation length $\xi_{SC}=3.04$.

\end{document}